\documentclass{aa}
\usepackage[varg]{txfonts}
\usepackage{graphicx}
\usepackage{natbib}
\usepackage{textcomp}
\begin{document}
\title{The Solar Cycle Variation of Topological Structures in the Global Solar Corona } 

\author{S. J. Platten \inst{1} \and C. E. Parnell\inst{1} \and A. L. Haynes\inst{1} \and E. R. Priest\inst{1} \and D. H. Mackay\inst{1}}
\institute{School of Mathematics \& Statistics, University of St Andrews, St Andrews, Fife, KY16 9SS, Scotland}

\abstract{The complicated distribution of magnetic flux across the solar photosphere results in a complex web of coronal magnetic field structures. To understand this complexity, the magnetic skeleton of the coronal field can be calculated. The skeleton highlights the (separatrix) surfaces that divide the field into topologically distinct regions, allowing open-field regions on the solar surface to be located. Furthermore, separatrix surfaces and their intersections with other separatrix surfaces (i.e., separators) are important likely energy release sites.}{The aim of this paper is to investigate, throughout the solar cycle, the nature of coronal magnetic-field topologies that arise under the potential-field source-surface approximation. In particular, we characterise the typical global fields at solar maximum and minimum.}{Global magnetic fields are extrapolated from observed Kitt Peak and SOLIS synoptic magnetograms, from Carrington rotations 1645 to 2144, using the potential-field source-surface model. This allows the variations in the coronal skeleton to be studied over three solar cycles.}{The main building blocks which make up magnetic fields are identified and classified according to the nature of their separatrix surfaces. The magnetic skeleton reveals that, at solar maximum, the global coronal field involves a multitude of topological structures at all latitudes criss-crossing throughout the atmosphere. Many open-field regions exist originating anywhere on the photosphere. At solar minimum, the coronal topology is heavily influenced by the solar magnetic dipole. A strong dipole results in a simple large-scale structure involving just two large polar open-field regions, but, at short radial distances between $\pm 60^\circ$ latitude, the small-scale topology is complex. If the solar magnetic dipole if weak, as in the recent minimum, then the low-latitude quiet-sun magnetic fields may be globally significant enough to create many disconnected open-field regions between $\pm 60^\circ$ latitude, in addition to the two polar open-field regions.}{}{}

\maketitle

\section{Introduction}
The coronal magnetic field of the Sun is constantly evolving due to both short-lived, but large-scale, events (such as flares and CMEs), as well as the long-term trends of the 11-year solar cycle (such as the number of active regions and the variation in polar-field strength \citep{hathaway10}). The emergence of flux combined with photospheric motions \citep{sheeley05} creates an intricate system of magnetic sources at the photospheric level which acts as the footprint of a highly complex magnetic field in the corona. 

Magnetic fields may be observed directly at the photospheric level, but, to understand the complexity of the coronal magnetic field, theoretical models are needed to extrapolate from photospheric data. Over the years a wide variety of techniques have been developed to model coronal magnetic fields based on the input of observed photospheric fields. These include the potential-field source-surface (PFSS) model (\citeauthor{schatten69} \citeyear{schatten69}; \citeauthor{altschuler69} \citeyear{altschuler69}), the current-sheet source-surface model \citep{zhao94}, non-linear force-free models (\citeauthor{vanballegooijen00} \citeyear{vanballegooijen00}; \citeauthor{mackay06} \citeyear{mackay06}; \citeauthor{wiegelmann07} \citeyear{wiegelmann07}; \citeauthor{contopoulos11} \citeyear{contopoulos11}; \citeauthor{yeates12} \citeyear{yeates12}), magnetohydrostatic models  \citep{neukirch95}  and finally full MHD models (\citeauthor{riley06} \citeyear{riley06}; \citeauthor{devore08} \citeyear{devore08}; \citeauthor{lionello09} \citeyear{lionello09}; \citeauthor{downs10} \citeyear{downs10}; \citeauthor{feng12} \citeyear{feng12}). A complete review of these models is given in Section 3 of \citeauthor{mackayyeates_livrev} (\citeyear{mackayyeates_livrev}). 

No matter which type of coronal model is applied, magnetic topology has to be determined in order to understand the complex structure of the Sun's global field. Magnetic topology describes the connectivity of field lines (e.g., \citeauthor{longcope05} \citeyear{longcope05}) where the primary way we investigate the topology is by constructing the \emph{magnetic skeleton} (e.g., \citeauthor{priest_etal96} \citeyear{priest_etal96}). The magnetic skeleton consists of the null points, separatrix surfaces, spines and separators within the field. 

In three dimensions, a null point (location where $\textbf{B}=\textbf{0}$) produces two important field-line structures: a pair of \emph{spine lines} and a \emph{separatrix surface} (e.g., \citeauthor{cowley73} \citeyear{cowley73}; \citeauthor{LauFinn90} \citeyear{LauFinn90}; \citeauthor{parnell96} \citeyear{parnell96}; \citeauthor{priest_titov96} \citeyear{priest_titov96}). 
The spine lines are special field lines that are directed into (out of) a positive (negative) null point. The \emph{separatrix surface} is formed from an infinite number of field lines that are directed out from (into) a positive (negative) null. In addition to separatrix surfaces from null points, separatrix surfaces can also be traced from \emph{bald patches} which are segments of polarity inversion lines on the photosphere where coronal field lines touch the surface and are concave upwards  \citep{titov93,titovetal2011}. Finally, the intersection of two separatrix surfaces from opposite-polarity nulls forms a special field line called a \emph{separator} (e.g., \citeauthor{LauFinn90} \citeyear{LauFinn90}). This field line actually connects the two nulls and is a generic structure that is maintained even under slight perturbations of the field, since the two surfaces will remain intersecting even if they are moved by a small amount.

PFSS models combined with observational magnetogram data have been used to calculate the amount of open flux from the Sun (e.g., \citeauthor{wang02} \citeyear{wang02}). Such estimates have been compared with observations of the total interplanetary magnetic field at 1 AU (e.g., \citeauthor{wang02} \citeyear{wang02}; \citeauthor{riley07} \citeyear{riley07}). The amount of open flux, its latitudinal distribution and area variation with height above the photosphere, is associated with the mechanisms for accelerating the solar wind. For instance, \citeauthor{wang90} (\citeyear{wang90}) used a PFSS model to estimate the solar wind speed calculated from flux-tube expansions of the open-field regions. However, recently the importance of global coronal separatrix surfaces and quasi-separatrix layers (QSLs) for the solar wind has been described by \citeauthor{antiochosetal2011} (\citeyear{antiochosetal2011}) and \citeauthor{crooker12} (\citeyear{crooker12}). They suggest that the slow solar wind can be heated and accelerated by a web of coronal separatrix surfaces and quasi-separatrix layers which they call the ``S-web''. 

Topological magnetic features have been shown to play a key role in solar flares and coronal mass ejections (CMEs). For instance, coronal null points are central to the breakout model for CMEs \citep{antiochos_breakout} and both the Bastille day flare of 1998 \citep{aulanier00} and the flare occurring in AR10191 on 16th November 2002 \citep{masson09} can be explained by a model involving reconnection at a coronal null. Reconnection at QSLs and separatrix surfaces (from null points and bald patches) have also been shown to be associated with dynamic coronal events (e.g., \citeauthor{aulanier07} \citeyear{aulanier07}). \citeauthor{parnell10b} (\citeyear{parnell10b}) demonstrated that the interaction of an emerging flux region with overlying field takes place through reconnection at multiple separators. 

Although PFSS models have been around for a long time, the topology of the fields computed by them has never really been analysed. It has been found that, even under the potential approximation, the field-line mapping and connectivity patterns in the corona can become very complex. So simply tracing from the polarity-inversion line that exists on the source surface does not reveal all the topological detail of the coronal field below. For example, the presence of a parasitic polarity can make open-field regions (coronal holes) of the same polarity become completely disconnected from one another, in the sense that they no longer have any flux linking them \citep{titovetal2011}. To date no systematic study of the magnetic skeleton of the global corona has been carried out over solar cycle time-scales. Here, we undertake such a study. First, though, we mention two studies that looked at the number of null points in different solar cycles.

\citeauthor{longcope09} (\citeyear{longcope09}) made spectral estimates of the numbers of coronal null points above local regions of quiet Sun from 562 SoHO/MDI magnetograms spanning two solar cycles. These estimates are thought to be reasonably accurate in determining the numbers of nulls 1.5 Mm above the photosphere due to small-scale network and intra-network features. However, the study did not consider any active-region fields since the area covered by the photospheric magnetograms was too small to model the coronal field above accurately.  

\citeauthor{cook09} (\citeyear{cook09}) determined the numbers of null points in the global corona over 2 solar cycles with a PFSS model using magnetic-flux-transport simulations to determine the time evolution of the radial magnetic field at the photosphere. From their coronal fields they found that the number of null points varied in phase with the solar cycle. However, the simulated magnetograms used by \citeauthor{cook09} (\citeyear{cook09}) only account for the emergence of active regions and do not consider any smaller-scale fields. Thus, their model is not suitable for estimating the numbers of nulls during solar minimum. 

Here, we consider potential field source surface extrapolations from observed synoptic magnetograms of significantly higher resolution than those simulated by \citeauthor{cook09} (\citeyear{cook09}). \citeauthor{cook09} (\citeyear{cook09}) were restricted in their study because the surface diffusion term included in their flux transport model smoothed out small-scale variations. Thus, our simulations are likely to reveal more null points at both solar maximum (if they exist) and solar minimum. Furthermore, since our domain encompasses the whole solar corona, the magnetic fields we extrapolate do not have as detailed a local structure as those determined by \citeauthor{longcope09} (\citeyear{longcope09}) and so we will miss the multitude of coronal nulls they found, that reside low down in the corona. 

In this paper, we study the magnetic topology of {\bf 37} years worth of coronal magnetic fields constructed using a PFSS model (e.g., \citeauthor{vanBallCartPriest} \citeyear{vanBallCartPriest}; \citeauthor{mackayyeates_livrev} \citeyear{mackayyeates_livrev}) from synoptic magnetograms taken by Kitt Peak and the Solar Optical Long-term Investigations of the Sun telescope (SOLIS). After constructing the PFSS field, the magnetic skeleton is determined by applying the null-point finding code (\citeauthor{nullfinder} \citeyear{nullfinder}) and the separatrix-surface finding code (\citeauthor{ssfind10} \citeyear{ssfind10}), as well as a code to identify the location of bald patches and bald-patch separatrix surfaces. This allows us to examine the types of topological structure seen at solar maximum and solar minimum. The main building blocks of the global coronal fields are identified and characterised according to the nature of their separatrix surfaces. 

The paper is structured as follows. In Section 2, we describe the data used in our study and present the details of the PFSS models we extrapolate. In Section 3, the basic topological building blocks that arise are illustrated. Next, in Section 4, typical examples are used to explain the complex nature of the coronal field at cycle maximum and both types of cycle minimum (e.g., with a strong and a weak dipolar field). Then, in Section 5, the variation in the prevalence and location of topological features are considered over three solar cycles. Finally, the conclusions are presented in Section 6.

\section{Synoptic Data and the PFSS Model}
In order to understand the complex nature of the global corona a realistic field distribution for the whole Sun must be specified at the level of the photosphere. However, as only part of the photospheric magnetic field may be observed at any one time, we approximate the distribution at any instance by applying synoptic data. The synoptic data used here comes from the National Solar Observatory at Kitt-Peak in Arizona, USA. Synoptic magnetograms from Carrington rotation 1645 to 2007 (17th August 1976 to 29th August 2003) were taken with the vacuum telescope, while, for Carrington rotations 2007 onwards, they were taken with the Solar Optical Long-term Investigations of the Sun (SOLIS) telescope. The most recent synoptic map used in this study is from Carrington rotation 2144 which began on the 21st November 2013. The vacuum telescope observations have a resolution of 360 pixels in longitude and 180 pixels in equal steps of sine latitude. Two resolutions of SOLIS synoptic magnetogram data are available. The lower resolution data have the same resolution as the vacuum telescope observations so they are used here for continuity of studying long-term trends over many solar cycles. Magnetogram data from the Wilcox Solar Observatory (WSO) and the Mount Wilson Observatory both cover similar long time periods as the Kitt-Peak data used here. We chose to use the latter because (i) it has a significantly higher resolution that the WSO data and (ii) much better high-latitude coverage than the Mount-Wilson data. Both of these factors are important in determining the global topology, especially during periods of solar minima. 

In our potential-field model, these synoptic magnetogram maps of the radial component of the magnetic field are taken as the lower boundary which lies at the solar photosphere (1 $R_\odot$). For the upper boundary the field is assumed to have only a radial component. This upper boundary is known as the \emph{source surface} and is assumed to lie at a radial distance of 2.5$R_\odot$ based on eclipse and coronograph observations (see \citeauthor{mackayyeates_livrev} \citeyear{mackayyeates_livrev}).

In order to use the magnetograms as a lower boundary condition for the PFSS model their total net flux must be corrected to zero. This is done by calculating the total flux imbalance and correcting for it evenly across all cells of the magnetogram. Then, to remove any steep gradients, the magnetograms are smoothed using a Gaussian filter with a width of 2 degrees.

These corrected magnetograms form the lower boundary for the PFSS calculation code \citep{vanBallCartPriest}. This code solves Laplace's equation in 3D using spherical harmonics. The PFSS extrapolation is made from summing up these harmonics. To get a perfect extrapolation an infinite number of harmonics would need to be included. However as this is not practical a finite number of harmonics are used. Using the harmonic model number $l=81$ enables us to resolve 162 polarity reversals in $B_r$ around the longitudinal direction at the photosphere. By testing different numbers of harmonics, we find that we may include harmonics up to 81 to give us the highest resolution from the data without introducing spherical ringing. We choose a resulting 3D magnetic field on a grid with a resolution of 329 grid-points in longitude, 165 grid-points in latitude and 48 grid-points in radius between 1.0$R_\odot$ and 2.5$_\odot$. The resolution in longitude and latitude is slightly less than that of the input data. The grid spacing is even in latitude and longitude and increases exponentially with radius so the resolution is higher closer to the photosphere. The lowest grid point is at the solar surface, 1$R_\odot$, and the next lowest grid point is at 1.019$R_\odot$.

\section{Topological Features of the Solar Corona}
\label{topol_features}
As already mentioned, the topology of magnetic fields is investigated by constructing their magnetic skeletons. The magnetic null points are found by applying the null-point finding code (\citeauthor{nullfinder} \citeyear{nullfinder}). At the outer boundary, the non-radial components of the magnetic field drop to zero. This means that anywhere the radial component drops to zero forms a null line at the outer boundary which can be thought of as a line of infinitely many null points. This null line on the source surface, which marks the base of the \emph{heliospheric current sheet} (HCS), is located by identifying the line along which the radial component of the magnetic field, $\textbf{B}=(B_r,0,0)$, switches direction from outwards (positive) to inwards (negative). 
The separatrix surfaces, spines and separators associated with these three-dimensional null points and null line are found using the separatrix-surface finding code (\citeauthor{ssfind10} \citeyear{ssfind10}). Additionally, bald patches on the photosphere and their associated bald-patch separatrix surfaces are located in a similar manner to the HCS null line and associated features. 

\begin{figure}
\centering{\includegraphics[width=0.8\linewidth]{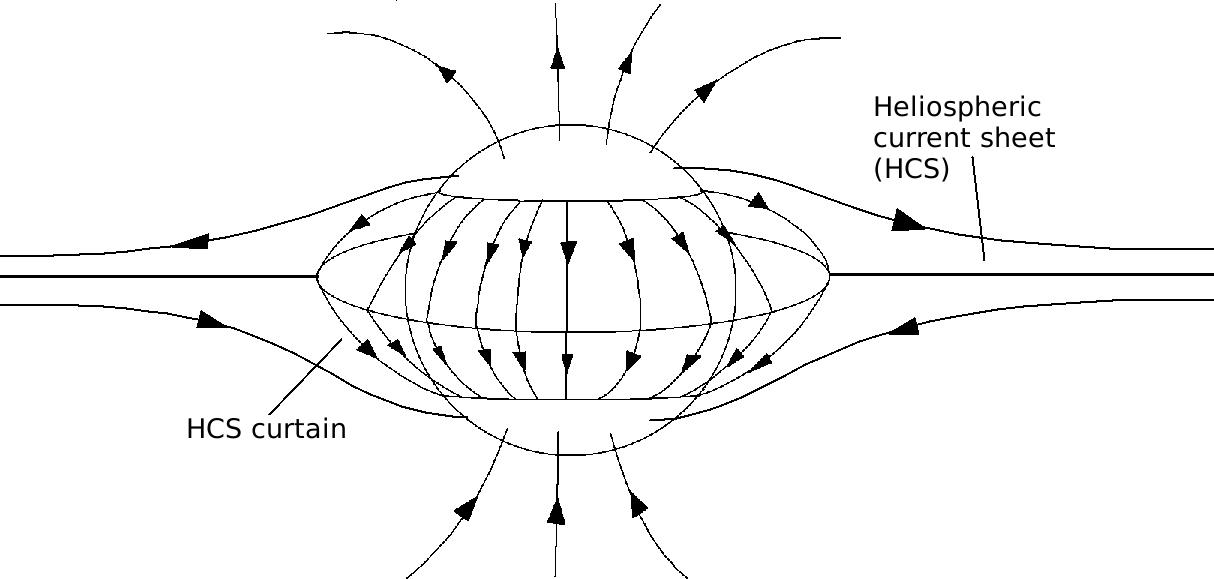}}
\caption{Cartoon of field lines traced from the base of the heliospheric current sheet, showing the heliospheric current sheet (HCS) and the HCS curtain.}
\label{hcs_cartoon}
\end{figure}
There are several types of topological feature, some of which are specific to a global model of the corona, which are found when constructing the magnetic skeletons. Figure \ref{hcs_cartoon} shows a sketch of field lines traced down from the HCS null line. These field lines produce surfaces that separate closed field from open field \citep{titovetal2011}. We refer to these surfaces as the \emph{heliospheric current sheet (HCS) curtains}. 

The structures created by the separatrix surfaces associated with the coronal null points can be characterised as follows. If the separatrix surface extends up to the source surface at 2.5$R_\odot$ then it will form (part of) a \emph{separatrix curtain} which may be \emph{closed} or \emph{open}. \emph{Closed separatrix curtains} are ``walls'' or ``surfaces'' made up of one of more connected separatrix surface from null points of the same sign. They are bounded on both sides by the HCS curtain. These structures are known to be associated with pseudostreamers. \emph{open separatrix curtains}, on the other hand, are the same, except they are only bounded on at most one side by the HCS curtain. The open sides of these curtains are bounded by spines from nulls of opposite sign to the nulls that produce the separatrix curtain. It is not known whether these structures may form pseudostreamers or not. Illustrations of these different types of separatrix curtains are shown below in Figures~\ref{doubledome3D},~\ref{cave3D}~and~\ref{tunnel3D}.

If the separatrix surfaces do not reach the source surface they form either a \emph{separatrix dome, cave} or \emph{tunnel}. Each of these structures is described and illustrated in Sections \ref{sep_dome_sec} and \ref{sep_cave_sec} using an idealised configuration of magnetic field at the photosphere. 

\begin{figure}
\centering{\centerline{\Large \bf \hspace{-0.5 \linewidth}{\tiny(a)}}
\includegraphics[width=0.95\linewidth]{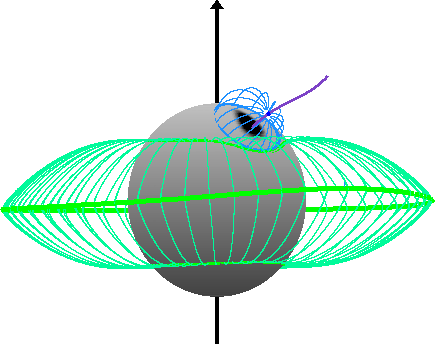}
 \centerline{\Large \bf \hspace{-0.5 \linewidth}{\tiny(b)}}
\includegraphics[width=0.95\linewidth]{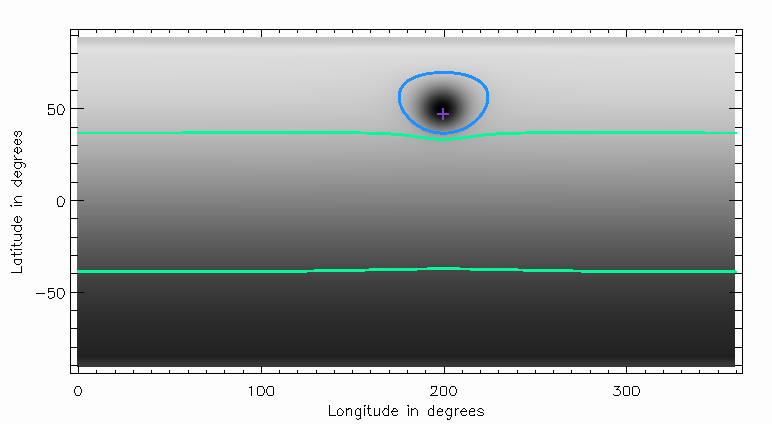}}
\caption{Example of the topology of a \emph{separatrix dome} from one separatrix surface. (a) 3D view of the topology consisting of a negative null (blue dot), its spine lines (thick purple lines) and separatrix surface field lines which arch down to the photosphere forming a dome (thin blue lines). The HCS null line (thick green line) and HCS curtain field lines (thin green lines) are also shown. (b) The magnetic field two grid points above the photosphere (grey-scale contour map) with the intersection of magnetic skeleton at this height over-plotted, including the separatrix dome (blue line), spine (purple cross) and HCS curtains (green lines).}
\label{dome3D}
\end{figure}
\subsection{Separatrix Domes}
\label{sep_dome_sec}
Separatrix domes form over parasitic polarity patches. They are made up of separatrix surfaces from one or more nulls. Figure \ref{dome3D}a shows the topology of a 3D global potential field with a HCS null line and HCS curtain (green) and also a separatrix dome made from a single separatrix surface (blue). In this example the dome is situated outside the HCS curtain so the spine (purple) extends out to the outer boundary. However, domes are frequently found inside the HCS curtain in which case both spines are connected to the photosphere. Figure \ref{dome3D}b shows the photospheric field and separatrix surfaces just above the lower boundary. The separatrix dome intersects with the photosphere in a closed curve that surrounds the parasitic polarity patch. In this case it is a negative patch embedded in the positive northern polar field, so a negative null forms over it.
\begin{figure}
\centering{
 \centerline{\Large \bf \hspace{-0.5 \linewidth}{\tiny(a)}}
\includegraphics[width=0.95\linewidth]{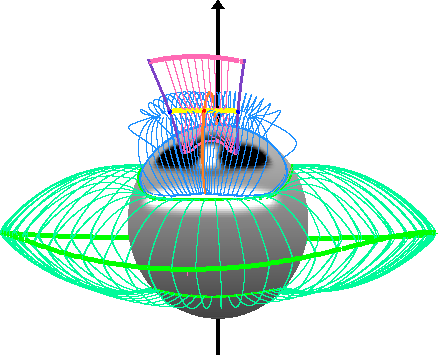}
\centerline{\Large \bf  \hspace{-0.5 \linewidth}{\tiny(b)}}
\includegraphics[width=0.95\linewidth]{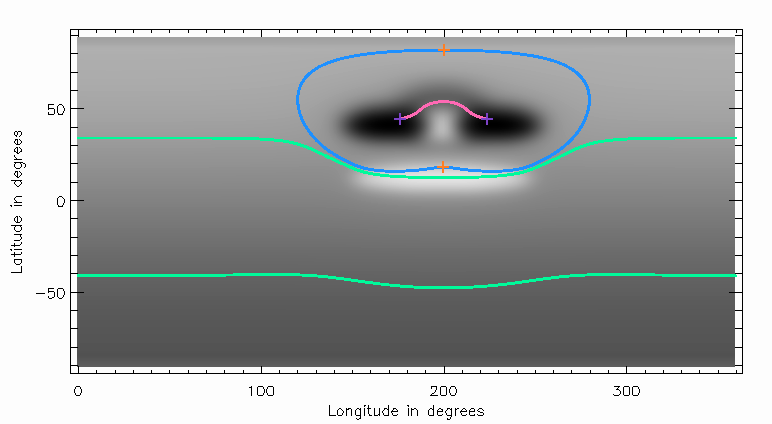}
}
\caption{Example of the magnetic topology of a \emph{separatrix dome} formed from two separatrix surfaces and an \emph{open separatrix curtain} which threads through the dome. (a) 3D image of the topology consisting of two negative nulls (blue dots), their spine lines (thick purple lines) and separatrix-surface field lines which arch down to the photosphere forming a dome (thin blue lines). A positive null (red dot), its spine lines (thick orange lines) and separatrix-surface field lines which form the open separatrix curtain (thin pink lines). The HCS null line (thick green line) and HCS curtain field lines (thin green lines) are also shown. (b) The  magnetic field two grid points above the photosphere (grey-scale contour plot) and the intersection of the skeleton features with this surface,  including the separatrix dome (blue line), spines (purple crosses) and HCS curtains (green lines). }
\label{doubledome3D}
\end{figure}

Figure \ref{doubledome3D} shows the case of a separatrix dome formed by two separatrix surfaces from nulls of the same sign, in this case they are negative nulls. These separatrix surfaces join along the spine of a positive null that lies between the two negative nulls. The separatrix surface from the positive null reaches down below to the photosphere and up above to the outer boundary of the model  forming an \emph{open separatrix curtain} which is bounded on either side by the spines of the negative nulls. The intersection of the separatrix surfaces with the photosphere can be seen in Figure \ref{doubledome3D}b, a separatrix dome made of two separatrix surfaces is characterised at the base by a closed loop intersected by spines with a segment of separatrix surface (open separatrix curtain) of opposite polarity inside the loop.

\subsection{Separatrix Caves and Tunnels}
\label{sep_cave_sec}
A separatrix dome that does not intersect with the photosphere on all sides, but is partially bounded by the spines of opposite-polarity nulls that do not bound any other separatrix surfaces, may form an ``open'' separatrix dome. There are two types of ``open'' separatrix domes known as \emph{caves} when they have one opening and \emph{tunnels} when they have two or more openings. 
\begin{figure}
\centering{\centerline{\Large \bf \hspace{-0.5 \linewidth}{\tiny(a)}}
\includegraphics[width=0.95\linewidth]{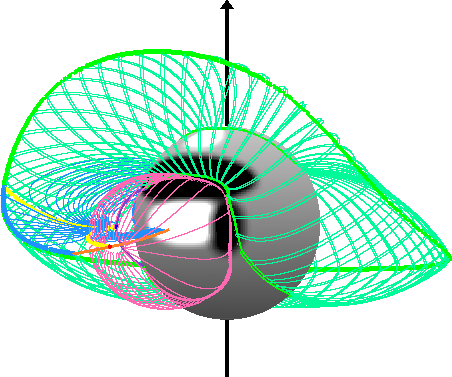}
 \centerline{\Large \bf \hspace{-0.5 \linewidth}{\tiny(b)}}
\includegraphics[width=0.95\linewidth]{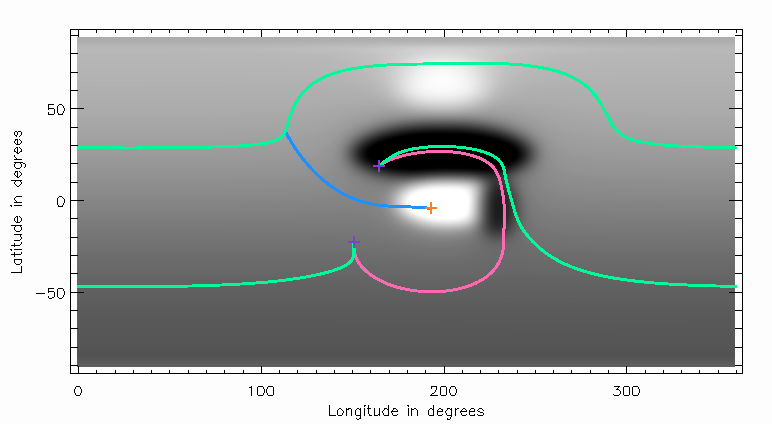}}
\caption{Example of the topology of a \emph{separatrix cave} formed from one separatrix surface and  a semi-open separatrix curtain. (a) 3D image of the magnetic topology. (b) The magnetic field two grid points above the photosphere (grey-scale contour map) and the intersection of the magnetic skeleton with this surface. Colour surfaces, lines and symbols as in Figure \ref{doubledome3D}}.
\label{cave3D}
\end{figure}

Figure \ref{cave3D} shows the topology of a separatrix cave. One side of a positive separatrix dome which is situated in a negative southern hemisphere field is bounded by the spines of a negative null that run along the edge of the HCS curtain. The separatrix surface from this negative null forms an open separatrix curtain which connects to the HCS at one end and at the other is bounded by the almost radial spine of the positive null. Since the separatrix surfaces from the two nulls intersect one another they form a separator linking the nulls, we refer to this type of separator as a \emph{null-null separator}. In this model there is also a separator formed from the intersection of the separatrix curtain with the HCS curtain  and connects a null point to the null line: we call this a \emph{null-HCS separator}.

The intersection of these features with the photosphere can be seen in Figure \ref{cave3D}b. The intersection of the separatrix cave with the photosphere is shown by the pink line. The mouth of the cave is bounded by negative spines and the cave opening is directed underneath the separatrix curtain.

\begin{figure}
\centering{
\centerline{\Large \bf     
      \hspace{-0.5 \linewidth}{\tiny(a)}} \hfill
\includegraphics[width=0.9\linewidth]{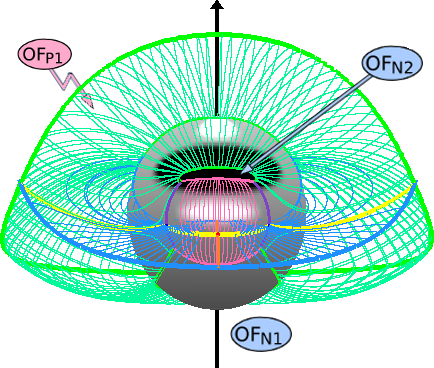}
\centerline{\Large \bf     
      \hspace{-0.5 \linewidth}{\tiny(b)} } \hfill
\includegraphics[width=0.9\linewidth]{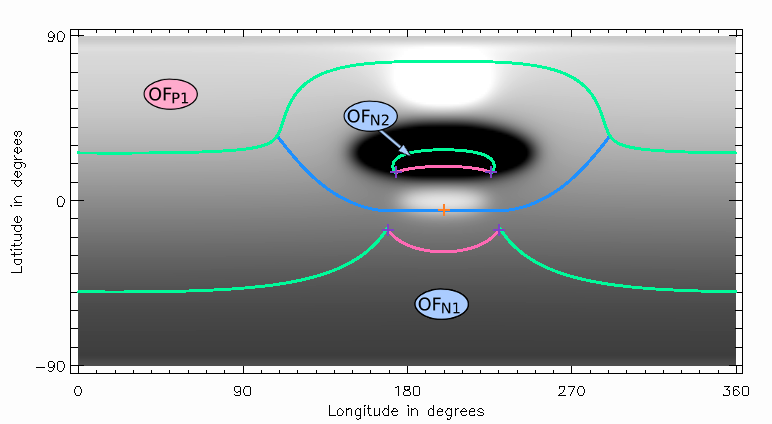}
\centerline{\Large \bf     
      \hspace{-0.5 \linewidth} {\tiny(c)}} \hfill
\includegraphics[width=0.9\linewidth]{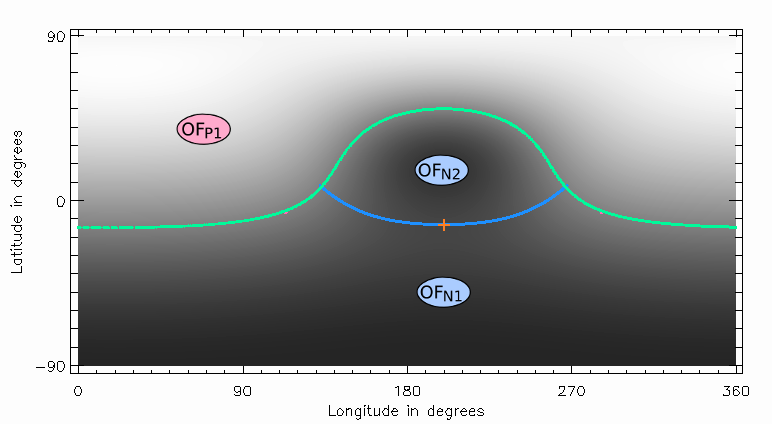}
}
\caption{Example of a \emph{separatrix tunnel} and \emph{closed separatrix curtain} topology. (a) 3D image of the magnetic topology. (b) The photospheric magnetic field and intersection of the magnetic skeleton with the photosphere. (c) Intersection of the magnetic skeleton with the source surface over-plotted on a contour plot of the source-surface radial field. Colouring and symbols as in Figure \ref{doubledome3D}.}
\label{tunnel3D}
\end{figure}

We refer to a separatrix dome with more than one opening as a \emph{separatrix tunnel}. Figure~\ref{tunnel3D} shows a field configuration that produces such a structure.
In this example the separatrix surface from a positive null residing in the weak negative southern hemisphere field is bounded by the spines from two negative nulls and intersects the photosphere in two segments (pink lines in Figure~\ref{tunnel3D}b), forming a separatrix tunnel. The separatrix surfaces from the two negative nulls are both bounded on one side by the spines of the positive null and on the other side by the HCS curtain, so together they form a \emph{closed separatrix curtain}. When such a curtain forms it creates a barrier between different open-field regions on the photosphere. In Figure~\ref{tunnel3D}b the small area of negative open field which has become disconnected from the southern negative coronal hole is labelled. 

The closed separatrix curtain intersects twice with the HCS curtain and forms two null-HCS separators. A pair of field lines in the separatrix surfaces of the central positive null connect to the two negative nulls, forming two null-null separators. So in total there are four separators present in this system.

Figure \ref{tunnel3D}c shows the intersection of the magnetic skeleton with the source surface and is over-plotted on contours of the radial field at this surface. The only magnetic skeleton features at the source surface are the HCS null line and the closed separatrix curtain which can be seen to be made up of two separatrix surfaces with the spine (orange cross) along its length indicating the join between the two surfaces. The three open-field regions present in this magnetic field are labelled ($OF_{P1}$, $OF_{N1}$ and $OF_{N2}$) in Figure~\ref{tunnel3D}. 

Note, that the sides and the mouths of a tunnel or cave may vary in size considerably and also there may be more than two entrances to a cave/tunnel. This means that separatrix-curtain structures may look more like complex bridges, cave systems or gazebo like canopies.
  
\section{Typical Global Topology at Solar Maximum and Solar Minimum} \label{max_min}
The topology of the PFSS models of the coronal field for every Carrington rotation with Kitt-Peak/SOLIS synoptic data over the last 37 years have been analysed. These show significant differences between the topology of the global coronal magnetic fields found at solar maximum and solar minimum.

\begin{figure}
\includegraphics[width=1\linewidth]{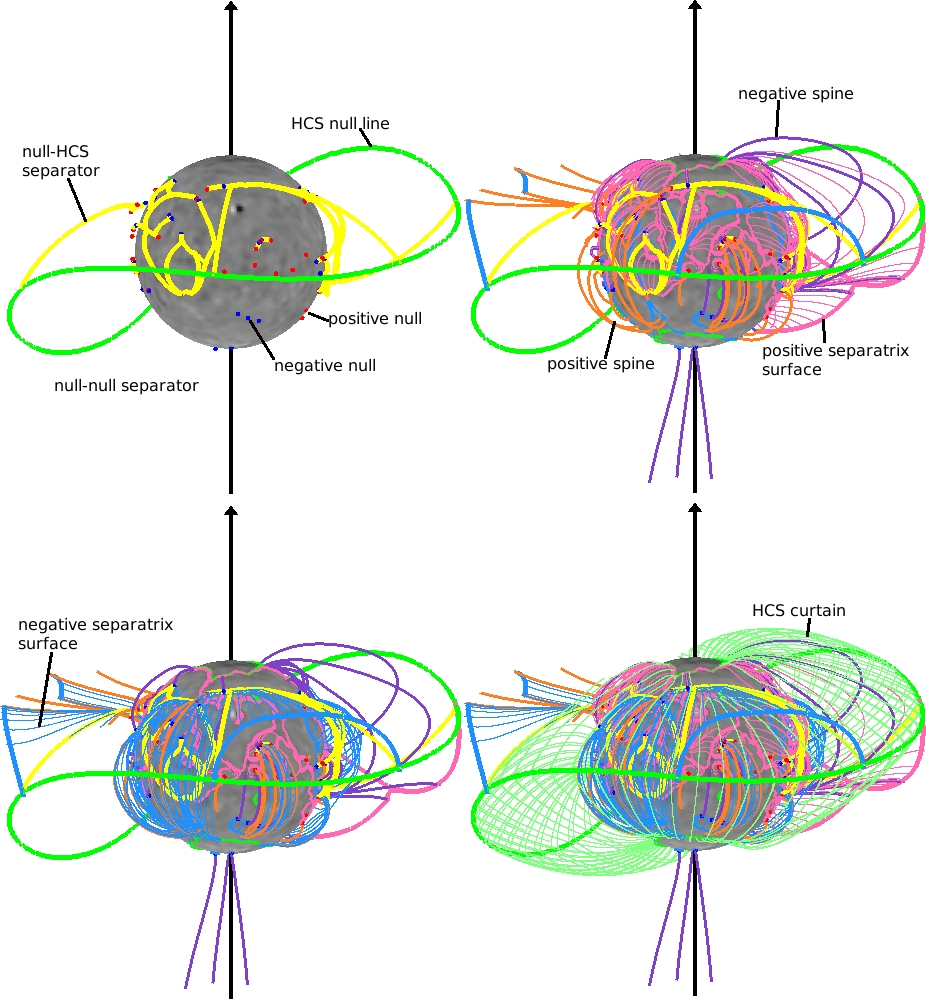}
\caption{The building blocks of global coronal topology. This example is Carrington rotation 2074 (August 2008, solar minimum cycle 23/24). The Sun is shaded with the radial component of the magnetic field at the photosphere. Red and blue dots represent positive and negative null points. Separators are thick yellow lines. Separatrix surfaces are traced from the nulls, as well as from the null-line at the base of the heliospheric current sheet (HCS) at the upper boundary of the model (green line) and from bald patches at the lower boundary of the model. The HCS curtain is represented by the bright green field lines, bald patch separatrix surfaces (if present) are represented by the olive green field lines, separatrix surfaces from positive/negative nulls are represented by pink/blue field lines. Spine lines from positive/negative nulls are orange/purple.}
\label{building_blocks}
\end{figure}

This section examines typical examples of the types of global topology formed by the potential coronal fields at solar minimum and solar maximum. Figure \ref{building_blocks} describes the scheme used to visualise the topology found using the techniques of \citeauthor{nullfinder} (\citeyear{nullfinder}, \citeyear{ssfind10}). Small red and blue dots represent positive and negative null points, respectively. The field lines in the separatrix surfaces traced from the nulls are pink for positive nulls and blue for negative nulls. Similarly the spine lines from positive and negative nulls are shown by thick orange and purple lines, respectively. Field lines in separatrix surfaces from bald patches are olive green. Separators, field lines that connect a pair of null points are represented by thick yellow lines. HCS curtain field lines are bright green. The surface of the Sun is shaded in grey-scale according to the radial component of the photospheric magnetic field and the axis of rotation is shown by the large black arrow. For ease of reference, each object and the colour/symbol used to represent it is listed in Table \ref{tab_col}.

\begin{table*}
\centering{
\begin{tabular}{|l|l|l|}
\hline
Topological feature & Colour & Symbol/Line style \\
\hline 
negative null point & blue & dot \\
positive null point & red & dot \\
negative separatrix surface field lines & light blue & thin lines \\
positive separatrix surface field lines & pink & thin lines \\
negative spines & purple & thick lines/crosses \\
positive spines & orange & thick lines/crosses \\
HCS null-line & green & thick line(s) \\
HCS curtain field lines & green & thin lines \\
bald-patch separatrix surface field lines & olive green & thin lines \\
separators & yellow & thick lines/stars \\
\hline
\end{tabular}}
\caption{List of the colours and symbols used to depict the elements of the 3D skeletons drawn in the 3D magnetic topology figures.}
\label{tab_col}
\end{table*}

\subsection{Typical Solar Minimum Topologies}
\label{sec:sol_min}
At solar minimum there are few or no active regions on the Sun and the magnetic field consists of the polar field (the solar magnetic dipole) and small-scale features. The resulting topology depends on the strength of the solar dipole, thus, giving two different kinds of solar minima: strong solar dipole minima and weak solar dipole minima.

\subsubsection{Strong Solar Dipole Minimum}
 If the solar dipole is strong, as in the minima seen between cycles 21 and 22 and between cycles 22 and 23, then few null points sit high in the solar atmosphere (i.e., above 70 Mm). The HCS null line lies near the equator and its curtains enclose most of the field at low latitudes (between $\pm50^\circ$) leaving large open-field regions at the poles, although additional open-field regions may be present due to closed separatrix curtains.

\begin{figure}
\includegraphics[width=0.9\linewidth]{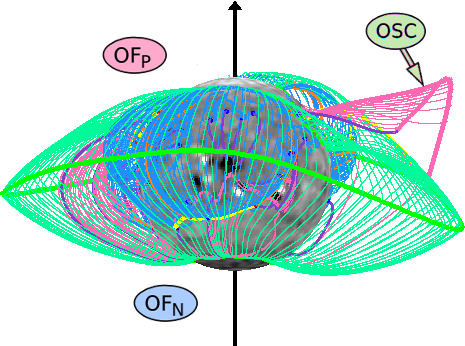}
\caption{Strong solar dipole minimum: Topology of PFSS extrapolation from Carrington rotation 1904. 
The notation of symbols is the same as that used in Figure~\ref{building_blocks} and listed in Table~\ref{tab_col}. The positive and negative open-field regions are labelled $OF_{P}$ and $OF_{N}$, respectively, and there is one open separatrix curtain present labelled OSC.}
\label{fig:PFSS_global_1904}
\end{figure}

\begin{figure}
\centerline{\Large \bf     
      \hspace{-0.9 \linewidth}{\tiny(a)}}
\includegraphics[width=0.9\linewidth]{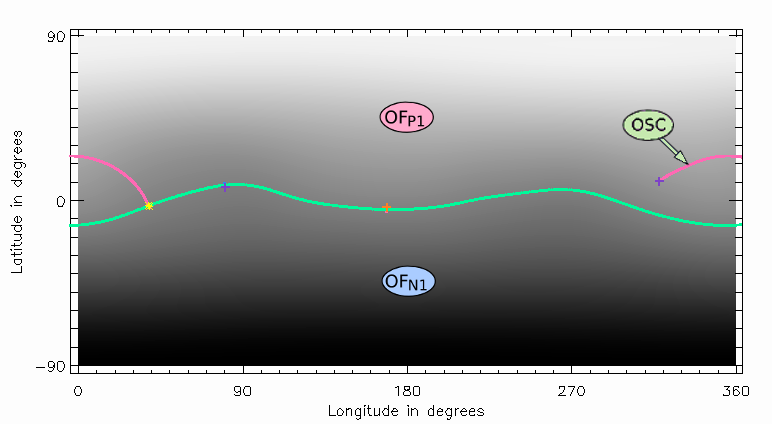}
\centerline{\Large \bf     
      \hspace{-0.9 \linewidth}{\tiny(b)}} 
\includegraphics[width=0.9\linewidth]{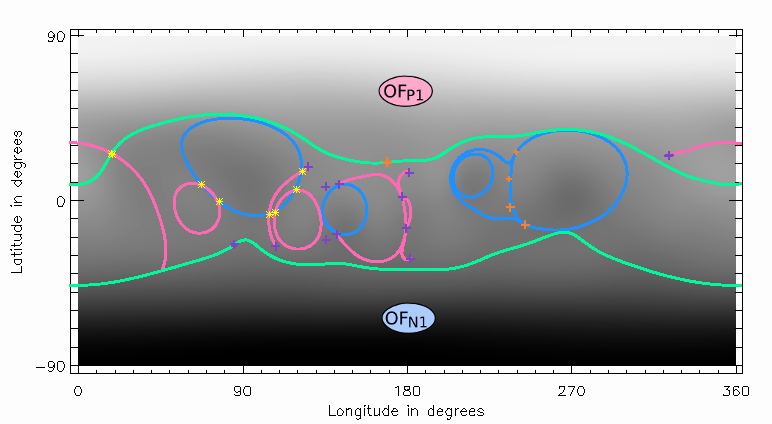}
\centerline{\Large \bf     
      \hspace{-0.9 \linewidth}{\tiny(c)}} 
\includegraphics[width=0.9\linewidth]{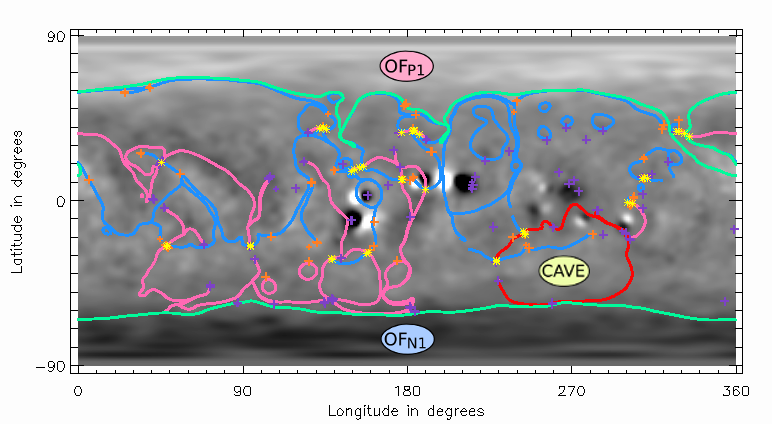}
\caption{Strong solar dipole minimum: Cuts in the separatrix surfaces at fixed radii of (a) $r=R_{ss}=2.5R_{\odot}$, (b) $r=1.44R_{\odot}$ and (c) $r=1.02R_{\odot}$. 
The magnetic topology features are coloured according to Figure~\ref{tunnel3D}b and Table~\ref{tab_col}. 
The polar open-field regions are the only open-field regions present and are labelled $OF_{P1}$ and $OF_{N1}$. The one open separatrix curtain is labelled OSC while the separatrix cave in (c) is labelled CAVE.}
\label{cuts_1904}
\end{figure}

Figure \ref{fig:PFSS_global_1904} shows an example of the topology of such a minimum from the PFSS model from Carrington rotation 1904 which is at the end of cycle 22. The HCS lies near to the equator and there is only one separatrix curtain present. This separatrix curtain (labelled OSC) is from a positive null point and is open; it is bounded by the HCS curtain and the spine of a negative null. Since it is bounded by a spine to which no other separatrix curtains connect, it does not form a disconnected open-field region. Only two open-field regions are present: one around the positive northern pole ($OF_{P}$), the other around the negative southern pole ($OF_N$). Almost all the features are enclosed within the HCS curtain, so to better examine the field, we take cuts through the magnetic skeleton at various radii in the solar atmosphere. We examine cuts at the outer boundary (2.5$R_\odot$), midway though the model (1.44$R_\odot$) and just above the photosphere (1.02$R_\odot$), Figure~\ref{cuts_1904}a,~\ref{cuts_1904}b~and~\ref{cuts_1904}c, respectively\footnote{Note that, for the cuts at the lower boundary, we take a cut at approximately 1.02$R_\odot$ as it is difficult to evaluate the intersections of the separatrix surfaces exactly on the lower boundary.}.

The only features that intersect with the outer boundary at 2.5$R_\odot$ (Figure \ref{cuts_1904}a) are the HCS null line and one positive open separatrix curtain. As already discussed, this separatrix curtain comes from a single null and is bounded on one side by a negative spine and on the other by the HCS curtain. 

Further down in the atmosphere more topological features are present. At a height of 1.44$R_\odot$ (Figure \ref{cuts_1904}b) there are thirteen separatrix surfaces intersecting with this height and producing seven separators. Near the solar surface (Figure \ref{cuts_1904}c) there are many more separatrix features present. One feature of note is the footprint of a separatrix cave (as shown in Figure \ref{cave3D}) made from the separatrix surfaces of several positive nulls intersecting with two separatrix surfaces from negative nulls and becoming bounded by their spines. The HCS curtain encloses almost all field between $\pm 60^{\circ}$ and so globally the field is very close to dipolar and the overall structure is very simple.
 
\begin{figure}
\includegraphics[width=0.9\linewidth]{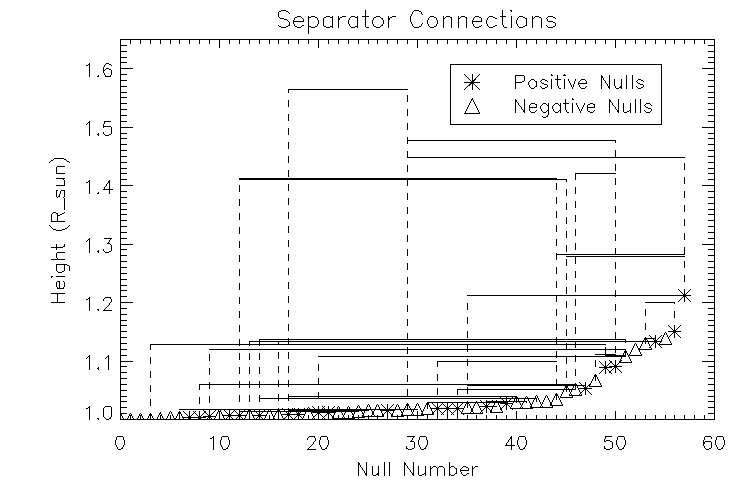}
\caption{Strong solar dipole minimum: Diagram of separator connections between null points. Positive and negative null points are represented by stars and triangles, respectively. If a separator links a pair of nulls, a line is drawn from the first null up to the maximum height reached by the separator and then along and down to the second null.}
\label{sep_con1904}
\end{figure}

From the cuts at fixed radii, we can see that there are many separators low down in the atmosphere, but few high up. Figure \ref{sep_con1904} summarises the heights reached by the nulls and the null-null separators. There are 58 coronal nulls present in this frame and 33 null-null separators. The highest null point sits at 1.21$R_\odot$, but the highest null-null separator reaches 1.56$R_\odot$. Although there are many separator connections most null-null separators are below 1.2$R_\odot$ which corresponds to a height of 139Mm. 

\subsubsection{Weak Solar Dipole Minimum}
On the other hand if the solar dipole is weak, such as in the recent solar minimum \citep{wang09}, then the global topology at solar minimum appears significantly different.
\begin{figure}
\includegraphics[width=0.9\linewidth]{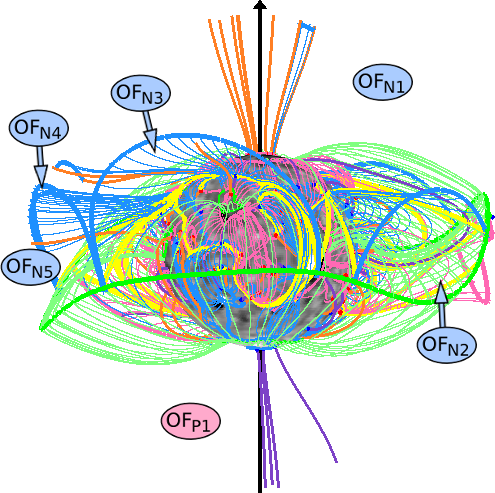}
\caption{Weak solar dipole minimum: Topology of PFSS extrapolation from Carrington rotation 2083. Features are coloured as in Figure \ref{fig:PFSS_global_1904}. There are 5 negative open-field regions visible labelled $OF_{N1-5}$ and one positive open-field region visible, labelled $OF_{P1}$}
\label{fig:PFSS_global_2083}
\end{figure}

Figure \ref{fig:PFSS_global_2083} shows the 3D topology at Carrington rotation 2083 beginning May 3rd 2009 (between cycles 23/24). The topology is much more complex than at CR 1904. The HCS is still fairly close to the equator (although it switches back and forth across the equator many times) and the polar regions are still open. Here, however, there are many more separatrix curtains, most of which are closed (connected to the HCS at both ends) and so form disconnected open-field regions. From the angle shown in Figure \ref{fig:PFSS_global_2083}, there are three negative open-field regions bounded to the south by the HCS curtain and to the north by closed negative separatrix curtains ($OF_{N2}$, $OF_{N3}$ and $OF_{N5}$). There is also one open-field region visible ($OF_{N4}$) which is bounded to the north and to the south by negative separatrix curtains.

\begin{figure}
\centering{
\centerline{\Large \bf     
      \hspace{-0.9 \linewidth}{\tiny(a)}} 
\includegraphics[width=0.9\linewidth]{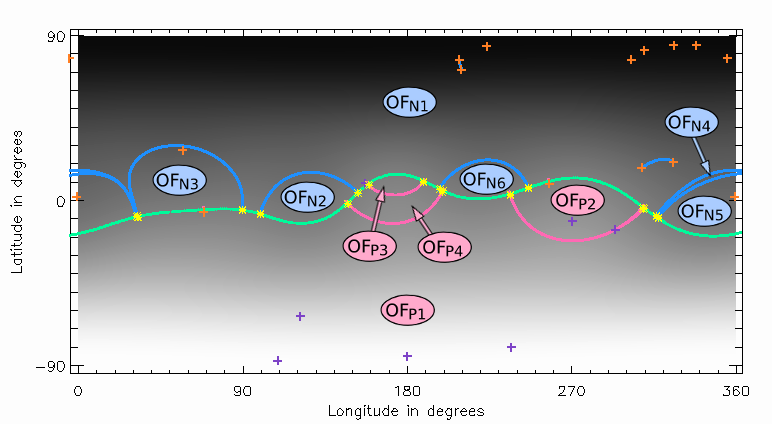}
\centerline{\Large \bf     
      \hspace{-0.9 \linewidth}{\tiny(b)} } \hfill
\includegraphics[width=0.9\linewidth]{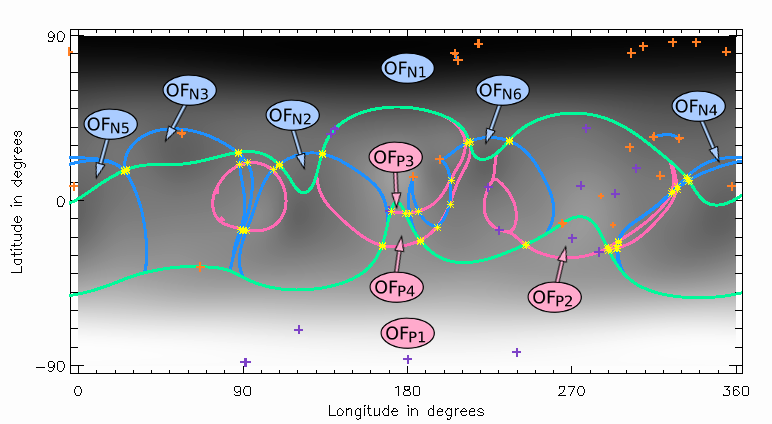}
\centerline{\Large \bf     
      \hspace{-0.9 \linewidth} {\tiny(c)}} \hfill
\includegraphics[width=0.9\linewidth]{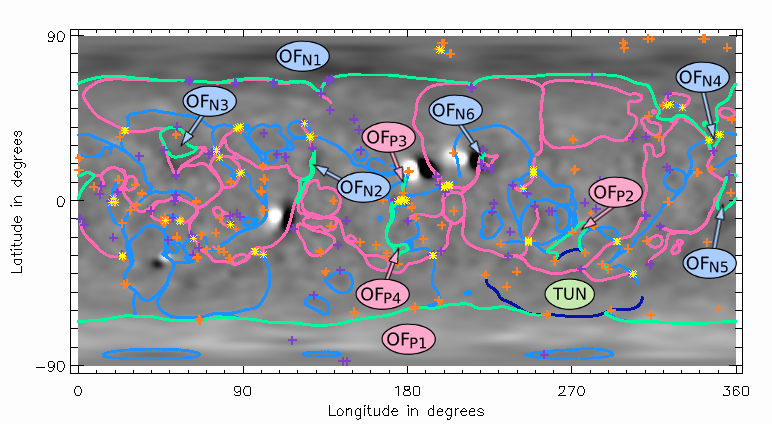}
}

\caption{Weak solar dipole minimum: Cuts through the magnetic skeleton of CR 2083 at fixed radii of (a) $r=R_{ss}=2.5R_{\odot}$, (b) $r=1.44R_{\odot}$ and (c) $r=1.02R_{\odot}$. Features are coloured as in Figure \ref{cuts_1904}. There are six negative open-field regions (labelled $OF_{N1-6}$) and four positive open-field regions (labelled $OF_{P1-4}$). The separatrix tunnel dividing $OF_{P1}$ and $OF_{P2}$ is labelled as TUN.}
\label{cuts_2083}
\end{figure}

Figure \ref{cuts_2083} shows cuts through the magnetic skeleton of Carrington rotation 2083 at fixed radii.
In the source surface cut (2.5$R_\odot$, Figure \ref{cuts_2083}a) the openings of the open-field regions have been labelled according to the scheme in Figure \ref{fig:PFSS_global_2083}. Additionally there is one more negative open-field region ($OF_{N6}$) and three more positive open-field regions ($OF_{P2}$, $OF_{P3}$, $OF_{P4}$) that are not really visible in Figure~\ref{fig:PFSS_global_2083}.

From the lower boundary cut (1.02$R_\odot$, Figure \ref{cuts_2083}c) we can identify features such as those described in Section \ref{topol_features}. At $270^\circ$ longitude and around $-50^\circ$ latitude at the lower boundary (Figure \ref{cuts_2083}c) there is a channel of negative separatrix surfaces with a positive separatrix surface intersecting it (labelled TUN). This is the footprint of a separatrix tunnel (see Figure \ref{tunnel3D}) and a closed separatrix curtain made of two separatrix surfaces. If we track this curtain to the top boundary it becomes the boundary for disconnected open-field region $OF_{P2}$. At this lower boundary, $OF_{P2}$ is separated from the polar open-field region $OF_{P1}$ by this separatrix tunnel. There is no clear field line path from the foot-points of $OF_{P1}$ to those of $OF_{P2}$ contrary to \citeauthor{antiochos2007} (\citeyear{antiochos2007}) who proposed the hypothesis that all open-field regions of the same polarity are connected at the photospheric level.

\begin{figure}
\includegraphics[width=0.9\linewidth]{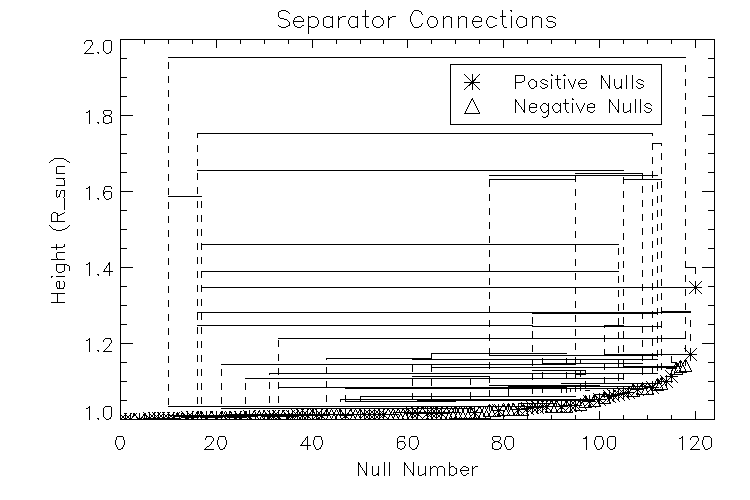}
\caption{Weak solar dipole minimum: Diagram of separator connections between null points for Carrington rotation 2083. Positive/negative null points are represented as stars/triangles. If there exists a separator between two nulls a line is drawn from the first null up to the maximum height reached by the separator and then along and down to the second null.}
\label{sep_con2083}
\end{figure}

From these cuts we can see that there are many separators present (yellow stars) at all heights. There are 73 null-null separators in total and 121 coronal nulls. The highest null is at a height of $1.35R_\odot$ and the highest null-null separator extends even higher (to 1.95$R_\odot$). Figure \ref{cuts_2083}c shows many null-null separators, however, most of these do not extend high into the atmosphere so few separators are visible in Figure \ref{cuts_2083}b and at the source surface (Figure \ref{cuts_2083}a) only separators arising from the intersection of the HCS curtain and separatrix surfaces from nulls occur. Figure \ref{sep_con2083} shows the heights and connections of the null-null separators. There are many connections, as in Carrington rotation 1904 (Figure \ref{sep_con1904}), and, although the null points still sit low in the atmosphere, the separators that connect them arch up high in the corona to much higher heights than when a strong solar dipole is present. This is due to the weak polar field allowing magnetic structures to expand. 

\subsection{Typical Solar Maximum Topology}
\label{sec:sol_max}
At solar maximum, many active regions are present on the solar surface with the global magnetic field dominated by these strong magnetic field features causing the HCS to become severely buckled \citep{titovetal2011} such that the photospheric field at the poles becomes closed. This allows the footprints of open-field regions to occur anywhere on the surface of the Sun, including equatorial regions.
\begin{figure}
\centering{\includegraphics[width=0.9\linewidth]{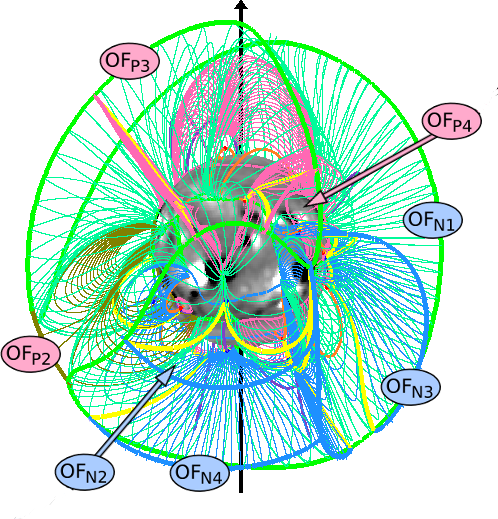}}
\caption{Solar maximum: Topology of potential field extrapolation from Carrington rotation 1957 (Solar Maximum). The same notation as Figure \ref{fig:PFSS_global_1904} is used. There are three positive polarity open-field regions visible ($OF_{P2}$, $OF_{P3}$, $OF_{P4}$) and four negative polarity open-field regions visible ($OF_{N1}$, $OF_{N2}$, $OF_{N3}$, $OF_{N4}$)} 
\label{fig:PFSS_global_1957}
\end{figure}

\begin{figure}
\centering{
\centerline{\Large \bf     
      \hspace{-0.9 \linewidth}{\tiny(a)} }
\includegraphics[width=0.9\linewidth]{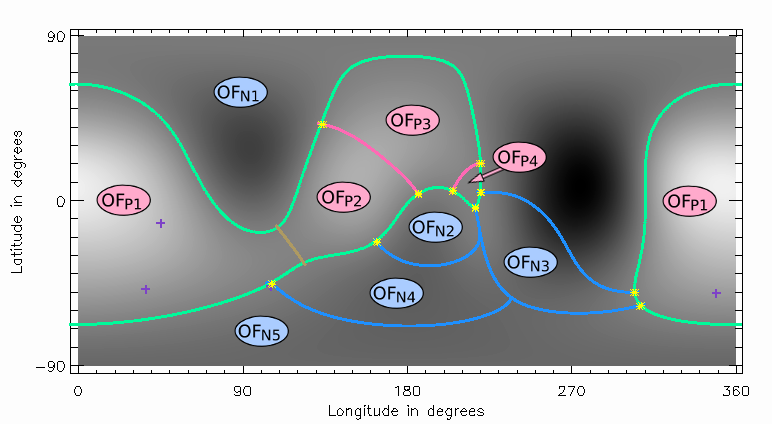}
\centerline{\Large \bf     
      \hspace{-0.9 \linewidth}{\tiny(b)} }
\includegraphics[width=0.9\linewidth]{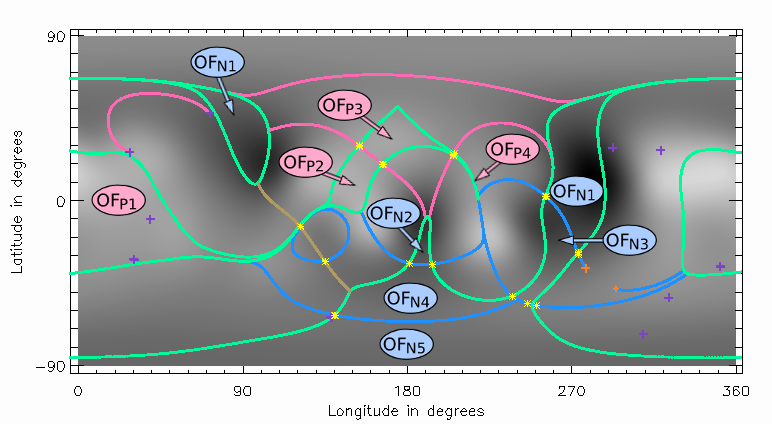}
\centerline{\Large \bf     
      \hspace{-0.9 \linewidth}{\tiny(c)} }
\includegraphics[width=0.9\linewidth]{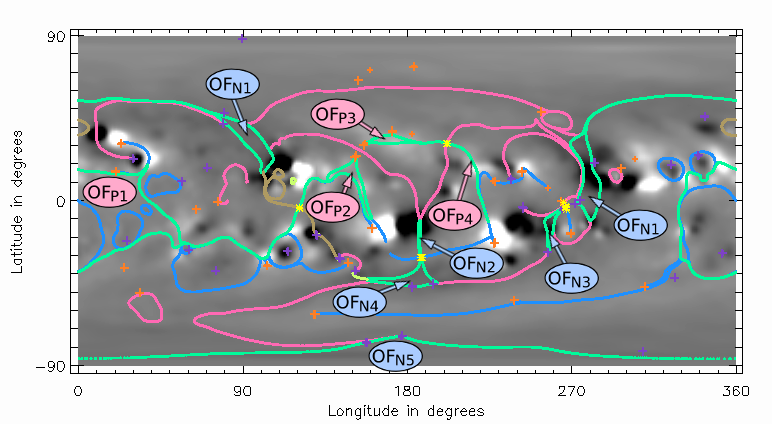}}
\caption{Solar maximum: Cuts at constant radii through the magnetic skeleton of Carrington rotation 1957 at (a) $r=R_{ss}=2.5R_\odot$, (b) $r=1.44R_{\odot}$ and (c) $r=1.02R_\odot$. The same notation as Figure \ref{cuts_1904} is used. There are nine different areas of open field labelled $OF_{N1-5}$ for the negative-polarity open-field regions and $OF_{P1-4}$ for the positive-polarity open-field regions.}
 \label{cuts_1957}
\end{figure}Figure \ref{fig:PFSS_global_1957} shows the 3D topology of a typical solar maximum field taken from CR1957, beginning on 5th December 1999. At the angle of the Sun shown, there are eight open-field regions: three with positive polarity ($OF_{P2-4}$) and four with negative polarity ($OF_{N1-4}$). There are seven separatrix curtains visible, six are from null points and one is from a bald patch (bounding the left of $OF_{P2}$). All the separatrix curtains are closed and divide the open field up into a total of nine regions. Figure \ref{cuts_1957}a shows a cut through the magnetic skeleton at the source surface (2.5$R_\odot$). The closed separatrix curtains found there form the boundaries of disconnected open-field regions. In addition to those seen in Figure \ref{fig:PFSS_global_1957}, there is another positive open-field region ($OF_{P1}$) that is bounded by the HCS curtain and a bald-patch separatrix surface. 

The HCS null-line (green line in Figure \ref{cuts_1957}a) does not follow close to the equator crossing all degrees longitude, as seen in the two solar minimum examples. Instead, it is absent between 220$^\circ$ and 300$^\circ$ longitude and crosses all other longitudes twice. This means that, rather than forming a wavy equatorial hoop, it forms a sort of squashed vertical closed loop that covers about 3/4 of the Sun. The warped nature of the HCS null-line is due to the relative strengths of the active-region magnetic fields in comparison to the polar field strength. In this instance the HCS is still only one loop but at some times during solar maximum the HCS may split to form multiple loops.

Using cuts at varying fixed radii, open-field regions can be tracked through the atmosphere to find their origins on the photosphere. Figure \ref{cuts_1957}(b) and (c) show cuts at $r=1.44R_\odot$ and $r=1.02R_\odot$, respectively. In these cuts through the magnetic skeleton, the regions of open field are much smaller and more distorted than they are at the source surface (Figure \ref{cuts_1957}a). At the photosphere the positive-polarity open-field region $OF_{P4}$ and the negative-polarity open-field region $OF_{N2}$ have shrunk so much that they are not visible when plotted at this resolution. 

Although in Figure \ref{cuts_1957}(c) few separatrix domes are visible at the photosphere, a number of these occur outside the HCS curtain and so reduce the region of open field within $OF_{P1}$. The limited occurrence of separatrix domes is because the topology is governed by large active regions, rather than the small-scale intermingling of flux that we see on the quiet Sun. The strong active region fields dominate over the weak polar fields, thus they tend to form large-scale topological structures that extend to the source surface.

\begin{figure}
\centering{\includegraphics[width=0.9\linewidth]{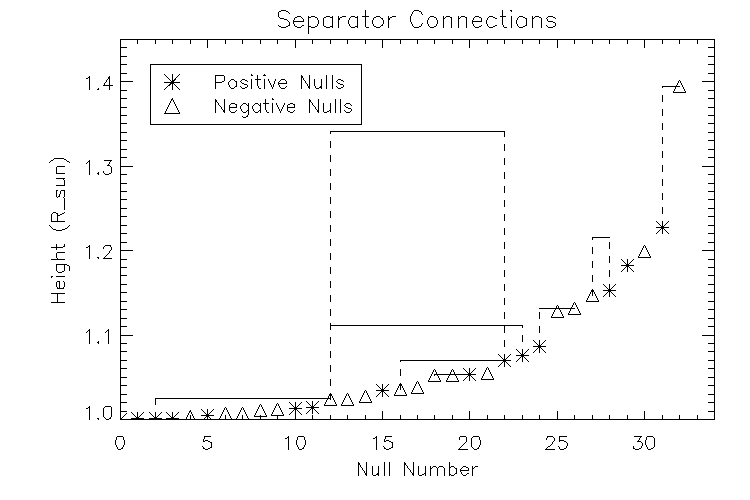}}
\caption{Solar maximum: Diagram of the separator connections between null points for Carrington rotation 1957. Positive and negative null points are represented as stars and triangles, respectively. If there exists a separator between two nulls a line is drawn from the first null up to the maximum height reached by the separator and then along and down to the second null.}
\label{sep_con1957}
\end{figure}

At solar maximum there are far fewer coronal null points than at minimum: in this frame 33 are present up to heights of 1.40$R_\odot$. There are also only a few null-null separators connecting these nulls (see Figure \ref{sep_con1957}). These null-null separators do not extend into the corona much higher than the null points themselves, unlike the solar minimum separators. In total, there are only 8 null-null separators present. This is likely to be for two reasons. Firstly, there are relatively few null points and these nulls are usually sufficiently far apart such that their separatrix surfaces do not interact. Secondly, the separatrix surfaces from these null points often intersect with the HCS curtains and so there are often null-HCS separators instead. The highest height reached by a null-null separator is 1.39$R_\odot$. Separators that reach higher in the corona at solar maximum are usually null-HCS separators.

\section{Solar Cycle Variation of Global Topological Properties}
In this section, the topological properties of 496 coronal potential-field source-surface extrapolations from every Carrington rotation for 37 years using the PFSS extrapolation code \citep{vanBallCartPriest} are studied\footnote{There are four Carrington rotations for which there is not a synoptic map: 2015, 2016, 2040 and 2041, and thus we cannot study their coronal topology.}. This allows a quantitative analysis over three solar cycles including the extended solar minimum at the end of cycle 23 and the beginning of cycle 24. The numbers of various topological features and also their positions show how the magnetic skeleton changes over the course of a solar cycle.
\subsection{Heliospheric Current Sheet Tilt}
As discussed in Section \ref{max_min}, the location and shape of the HCS changes greatly between solar minimum and solar maximum. For instance, it may form one or more closed null-line loops on the source surface \citep[e.g.,][]{wang14}. Also it is well established that the tilt angle (an indication of the latitudinal extent) of the HCS is positively correlated with the solar cycle (e.g., \citeauthor{hoeksema82} \citeyear{hoeksema82}; \citeauthor{smith01} \citeyear{smith01}; \citeauthor{riley02} \citeyear{riley02}; \citeauthor{jiang10} \citeyear{jiang10}). Here, we consider the tilt angle of the HCS from our the PFSS model using KP/SOLIS data and compare it against the HCS tilt angles found using the two Wilcox Solar Observatory (WSO) data sets (\url{http://wso.stanford.edu/Tilts.html}). The key difference between the KP/SOLIS and WSO data is that the KP/SOLIS data have higher resolution and, hence, higher harmonics are included in our PFSS extrapolations of the field.

\begin{figure}
\includegraphics[width=\linewidth]{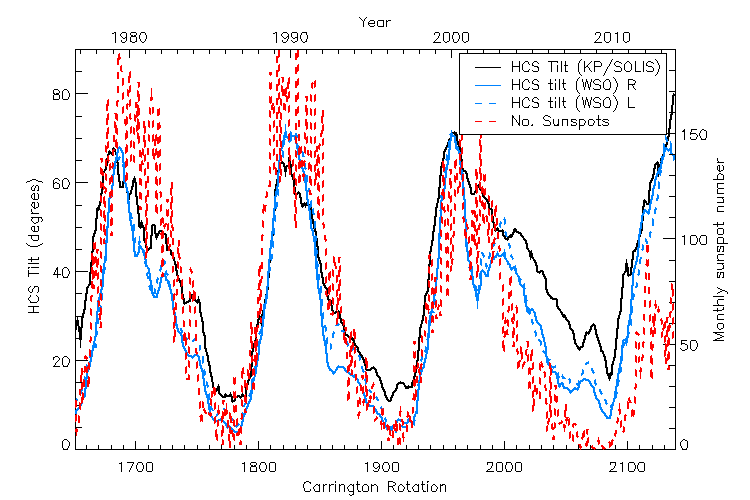}
\caption{ 12 CR running mean of the HCS tilt angle against time (solid line). The blue dashed and blue solid lines represent the HCS tilt angle calculated from the WSO data. The red dashed line represents the monthly sunspot number to indicate the solar maxima and minima.}
\label{hcs_tilt}
\end{figure}

Figure \ref{hcs_tilt} shows how the tilt angle of the HCS (i.e., the mean of the maximum displacements of the HCS in the northern and in the southern hemispheres in each Carrington rotation) varies with time for our KP/SOLIS model and the two WSO models. For comparison, the monthly sunspot number\footnote{Sunspot data comes from the Royal Greenwich Observatory up until 1977 and then afterwards from the US Air Force Solar Optical Observing Network: \url{http://solarscience.msfc.nasa.gov/greenwch/spot_num.txt}.} is also plotted (red dashed line) to indicate the solar cycles. 
 For all three models the tilt angle of the HCS agrees very well and follows the same trends as the sunspot number during cycles 21, 22 and the first half of cycle 23. However, after CR1970 for KP/SOLIS and CR1990 for WSO the HCS tilt results begin to diverge from the number of sunspots, with all three lines larger than might be expected in comparison to the two previous decline phases of the solar cycle, with the KP/SOLIS data predicting the largest tilt. The change from Kitt Peak to SOLIS data occurs at CR2007 which is after the divergence of the lines, so the change in behaviour is not caused by the changing data set. 

 Also, during cycle 24, all three data sets estimate the HCS tilt angle to be significantly larger than might be expected from the number of sunspots seen in cycle 24. Indeed, the HCS extends to latitudes similar to those seen during the previous two maxima, even though the sunspot numbers in cycle 24 are far lower. Such high tilt angles have been noted by \citeauthor{owens12} (\citeyear{owens12}) who studied proxies for the long-term HCS tilt and sunspot number back to the last Maunder minimum. They find that the modulation of the HCS tilt angle is not dependent on the sunspot number, but rather only on the phase of the solar cycle. 

From our studies of the complete topology throughout the global corona over several cycles, it seems that the increased tilt angle of the HCS is a result of the equatorial fields having a greater magnetic influence over the polar fields (i.e., the higher harmonics which describe the equatorial fields become comparable or greater in size than the lower harmonics). Of course, this effect is more evident in the KP/SOLIS data than in the WSO data, because the KP/SOLIS data have a higher resolution. During solar maxima the polar fields reverse, thus are weak, and strong fields from active regions disperse to dominate at all latitudes. So, due to this dispersal of the active-region fields the lowest-order harmonics are not particularly dominant. The equatorial quiet-sun fields during the solar minimum between cycles 23 and 24 were known to be weaker than in previous cycles \citep{thornton} however the polar fields at this time were also weak. This means the low-latitude quiet-sun fields had a greater global presence than in the previous two minima where the polar fields were strong. 

\begin{figure}
\centering{\includegraphics[width=0.9\linewidth]{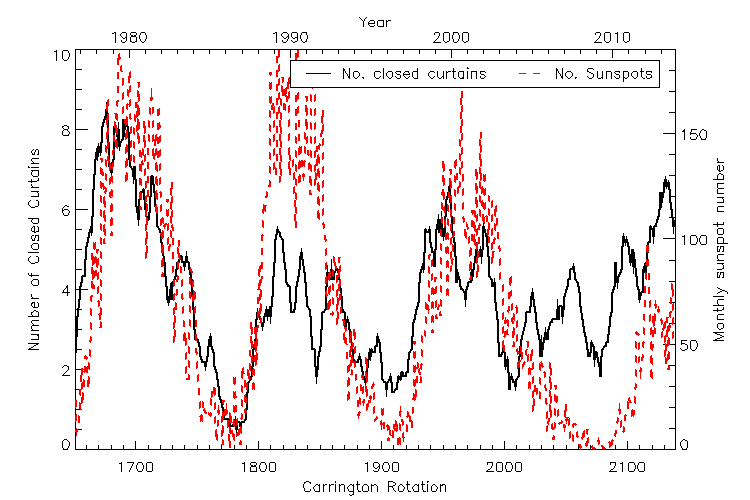}}
\caption{12 CR running mean of the number of closed separatrix curtains (or proxy for the number of disconnected open-field regions) against time. The {\bf red} dashed line shows the monthly sunspot number.}
\label{closed_curtains} 
\end{figure}
A measure of the amount of large-scale topological features (the influence of the higher-order harmonics/mixed polarity fields) can be made by looking at the number of closed separatrix curtains in each CR. As can be seen from Figure~\ref{closed_curtains}, closed separatrix curtains are abundant during solar maximum, when there are large active regions and weak polar fields and also during the weak global-dipole minimum between cycles 23 and 24. This graph therefore supports our explanation for the large HCS tilt. 

Furthermore, if a separatrix curtain is closed (bounded by the HCS at both ends) then it can form a disconnected open-field region at the photosphere (see Figure~\ref{tunnel3D} in Section~\ref{sep_cave_sec}). Hence, Figure~\ref{closed_curtains}, also suggests that the number of disconnected open-field regions will be greater when the HCS is more distorted (e.g. during solar maxima and weak solar dipolar minima). As discussed in \cite{wang07} the HCS null line is associated with helmet-streamers, whereas the closed-separatrix curtains are associated with pseudostreamers. Here, we clearly see that the number of pseudostreamers varies in phase with the solar cycle in agreement with \cite{owens13}. We also note that the closed separatrix surfaces (pseudostreamer) are believed to be sources of the slow solar wind (\citeauthor{wang12}, \citeyear{wang12}; \citeauthor{crooker12}, \citeyear{crooker12}), although others \citep{panasenco12} suggest they may also be sources of fast solar wind depending on geometrical nature of the associated open-field region. In agreement with our findings, \cite{owens13} found that pseudostreamers were more abundant during the declining phase of cycle 23 compared to cycles 21 and 22. They claim that this can largely be attributed to the extended length of the cycle 23. However, from our study of the topology of the global magnetic field we believe it is a result of the dipole fields not being replenished after the polar reversal at maximum in cycle 23. The weak solar dipole that persisted between cycles 23 and 24 lead to the mixed polarity low-latitude quiet-sun fields gaining a global presence (higher harmonics becoming important).



\subsection{Coronal Null Points}
\begin{figure}
\centering{\includegraphics[width=0.9\linewidth]{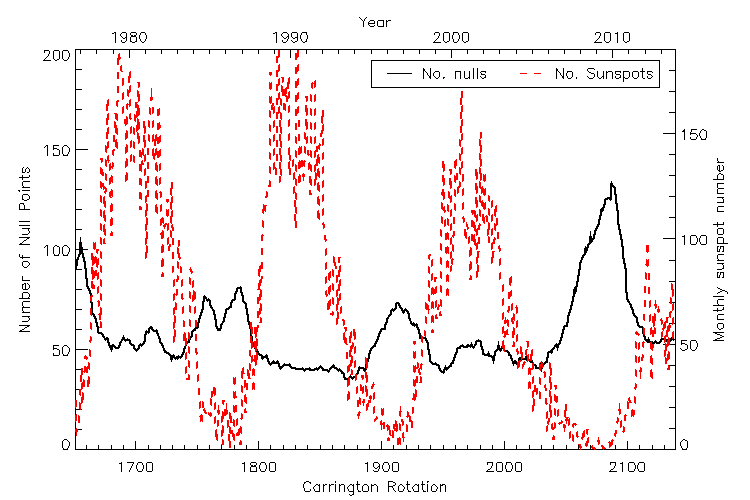}}
\caption{The variation in the number of nulls over time. The solid line shows a 12 CR running mean of the number of null points and the {\bf red} dashed line shows the monthly sunspot number.} 
\label{null_num}
\end{figure}The number of coronal null points varies greatly over a solar cycle as shown in Figure \ref{null_num}. From our PFSS results, at cycle minima, the total number of nulls in the corona is at its greatest. This is the converse to the result of \citeauthor{cook09} (\citeyear{cook09}) who found more nulls present during cycle maximum. This discrepancy arises since our data have both increased resolution and reduced smoothing of the photospheric field in comparison to the simulated data used by \citeauthor{cook09} (\citeyear{cook09}). This means that more small-scale photospheric features are resolved, hence, the extrapolated magnetic field is much more mixed on all scales, and so more null points are present, especially at solar minimum.

Interestingly, the most recent solar minimum has a much greater total number of nulls than the previous two minima considered. This is associated with the effects of the weak solar dipole and greater mixed small-scale field at low and high latitudes. More mixed polarity field means more null points in the corona. 

\begin{figure}
\centering{\includegraphics[width=0.8\linewidth]{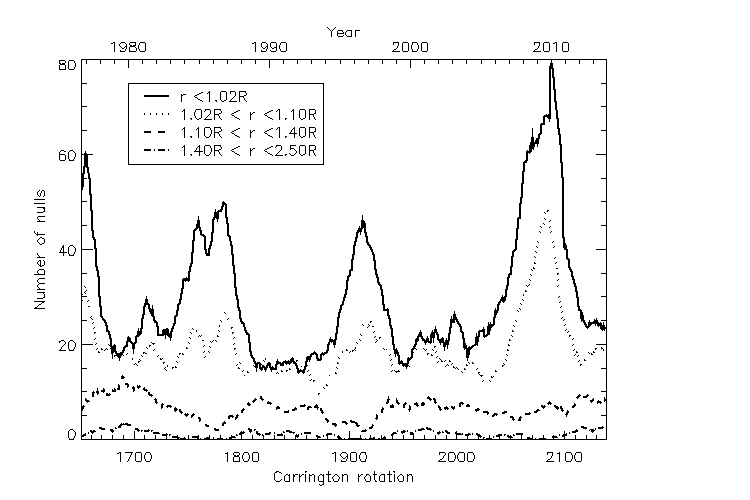}}
\caption{The variation in the height of null points against time. 12 CR running means of the number of nulls below 1.02 $R_{\odot}$ (solid black line), between 1.02$R_{\odot}$ and 1.1$R_{\odot}$ (dotted line), between 1.1$R_{\odot}$ and 1.4$R_{\odot}$ (dashed line) and above 1.4$R_{\odot}$ (dot-dashed line). } 
\label{null_height}
\end{figure}
Figure \ref{null_height} shows how the number of nulls that lie between different heights vary over the course of three solar cycles. The solid line shows the number of nulls at heights less than 1.02$R_{\odot}$ which corresponds to 14 Mm above the solar surface\footnote{It should be noted that the lower boundary is at 1$R_\odot$ and the next grid point is at $1.019R_\odot$ so the number of null points that can be found below 1.02$R_{\odot}$ will be limited by this resolution.}. At solar minimum, \citeauthor{longcope09} (\citeyear{longcope09}) estimates that at least 1 null point per 322 Mm$^2$ would be found below this height over mixed-polarity quiet-Sun regions. However, we find considerably fewer null points than this (about 100-150 times less) because of the resolution limits of our model. 

At solar minimum the nulls below 1.02$R_{\odot}$ (14 Mm) make up the majority of the nulls present in the corona. These nulls are associated with the intermingling of flux on the quiet Sun. This same process will account for any low down null points present at solar maximum. However, at solar maximum there are fewer of these low down nulls as more of the photospheric field becomes organised within large active regions. There are usually more high-altitude coronal nulls during solar maximum than minimum (as seen in the dashed and dot-dashed lines in Figure~\ref{null_height} corresponding to height ranges of 1.1$R_\odot$ to 1.4$R_\odot$ and 1.4$R_\odot$ to 2.5$R_\odot$). These high-altitude nulls are associated with active regions and are the sorts of nulls identified in the study by \citeauthor{cook09}~(\citeyear{cook09}). Contrary to this trend, the last solar minimum did not show a decrease in the number of nulls at heights between 1.1$R_\odot$ and 1.4$R_\odot$ (Figure~\ref{null_height}, dashed line). This is another consequence of the weak polar field strength during the last minimum enabling nulls associated with small-scale magnetic elements to occur higher in the atmosphere.

\begin{figure}
\centering{\includegraphics[width=0.99\linewidth]{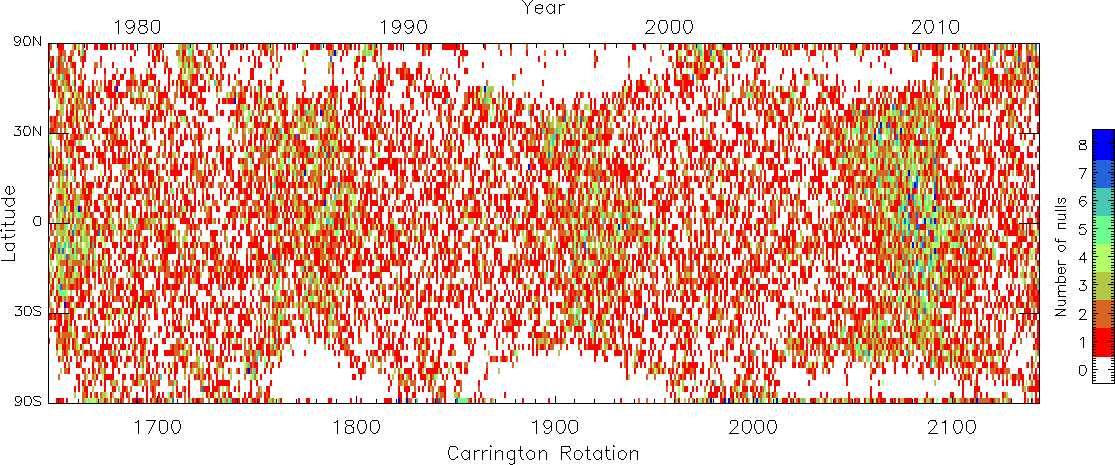}}
\caption{``Butterfly'' plot of the number of nulls present at each latitude against time. The colour represents the number of nulls in a particular latitudinal range for each synoptic map. The $y$ axis is scaled with the sine of the latitude so that every pixel represents the same area on the sun.}
 \label{null_lat}
\end{figure}
The latitude of null points, of course, vary over the cycle. Figure \ref{null_lat} shows a ``butterfly'' diagram of the total number of nulls at different latitudes for each Carrington rotation. At solar minimum there are more null points at low latitudes and few or no nulls at high latitudes\footnote{It should be noted that in the original synoptic maps the field above 70$^\circ$ is not well resolved: there are only 10 grid points above 70$^\circ$.}. This is caused by the polar fields being relatively strong and containing no large opposite-polarity features at solar minimum, so few nulls are detected over the polar regions. As the cycle progresses and meridional flows push residual patches of opposite-polarity active-region field to the poles leading to nulls being seen at high latitudes (as seen in Figure \ref{null_height} dotted line). During periods of solar maximum, when the mixing of polarities occurs at high latitudes, there is a good distribution of null points at all latitudes.

\begin{figure}
\includegraphics[width=1\linewidth]{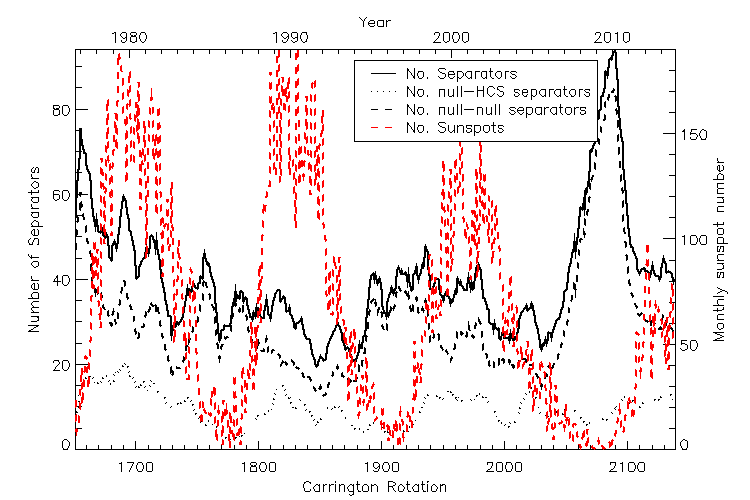}
\caption{The variation of the number of separators against time (solid black line). Also plotted is the monthly sunspot number (dashed red line), the number of null-null separators (black dashed line) and the number of null-HCS separators (black dotted line). All separator numbers are calculated using a 12 CR running mean.}
\label{sep_num}
\end{figure}
\begin{figure}
\includegraphics[width=1\linewidth]{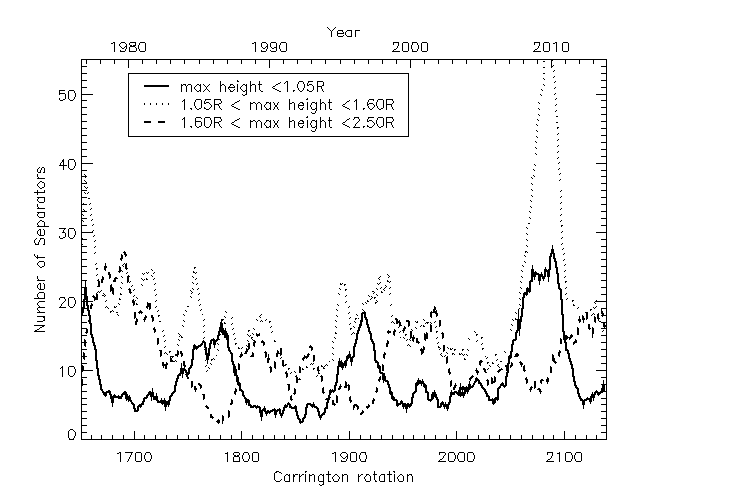}
\caption{The 12 Carrington rotation running mean time variation of the number of separators reaching maximum heights in the ranges: $R<1.05 R_{\odot}$ (solid line), $1.05 R_{\odot}\le R \le 1.6 R_{\odot}$ (dotted) and $1.6 R_{\odot}\le R \le 2.5 R_{\odot}$ (dashed).} 
\label{sep_height}
\end{figure}
\subsection{Separators}
 Separators are important sites for magnetic reconnection (e.g., \citeauthor{priest_titov96} \citeyear{priest_titov96}; \citeauthor{galsgaard00b} \citeyear{galsgaard00b}; \citeauthor{longcope05} \citeyear{longcope05}; \citeauthor{parnell10a} \citeyear{parnell10a}) and could be important for large energy release events such as solar flares and CMEs. These possible reconnection sites vary over the solar cycle in number, location and length. Figure \ref{sep_num} shows the variation in the number of separators over time. There is a negative correlation between the total number of separators present and the number of sunspots (the Pearson's correlation coefficient has been calculated as $-0.317$, which is around the 98.5\% significance level) as one might expect since the number of coronal null points is also out-of-phase with the solar cycle. However, more interesting conclusions can be drawn regarding the variation in the number of separators over the solar cycle by subdividing the separators into categories.

Firstly, separators may be categorised into {\it null-null separators}, that connect pairs of null points or {\it null-HCS separators} that connect a null point to the HCS null line. As one would expect, since the numbers of nulls are negatively correlated with the sunspot number, we find that the null-null separators are similarly negatively correlated (with a correlation coefficient of $-0.475$ which is at the 99.9\% significance level), see dashed line in Figure~\ref{sep_num}. On the other-hand the null-HCS separators show a positive correlation with the sunspot number (correlation coefficient of $0.513$, again at the 99.9\% significance level). This is due to the greater heights of nulls during maximum in comparison to minimum, as well as the increased global significance of the equatorial fields over the solar dipole at this time. For the same reasons the weak solar dipole in the most recent minimum (cycle 23/24) also results in an increased number of null-HCS separators as the globally significant low-latitude fields create more separatrix curtains connecting to the HCS null line than are present in the previous two minima.

Figure~\ref{sep_height} shows the variation in the number of separators reaching maximum heights in different ranges over the solar cycle. There is a clear cyclic variation in the number of low-lying separators which is out-of-phase with the solar cycle. This is in line with the results found in the case studies discussed in Section~\ref{max_min} and is also consistent with the out-of-phase cyclic variation found in the number of low-lying nulls. However, there is one interesting difference: during the last solar minimum the number of low lying nulls was unusually large leading as expected to many more low-lying separators, however, the greatest change is seen in the number of mid height separators. This fact, combined with the large total number of null-null separators during this minimum, is another indication of the global presence of the low-latitude small-scale fields due to the weak dipolar field.

At all times most separators have maximum heights between $1.05R_\odot$ and $1.6R_\odot$. The number of these mid-height separators fluctuates considerably, but with no clear cyclic variation following the solar cycle. However, there is a very definite peak in the number of these mid-height separators in the last solar minimum. This is consistent with the existence of a weak solar dipole field permitting the low latitude mixed polarity fields to expand and become more globally significant. Finally, in Figure~\ref{sep_height}, we see that the separators reaching maximum heights of more than $1.6R_\odot$ varies in phase with the solar cycle. This is due to the greater number of null-HCS separators that occur during solar maximum. 

\begin{figure}
\includegraphics[width=1\linewidth]{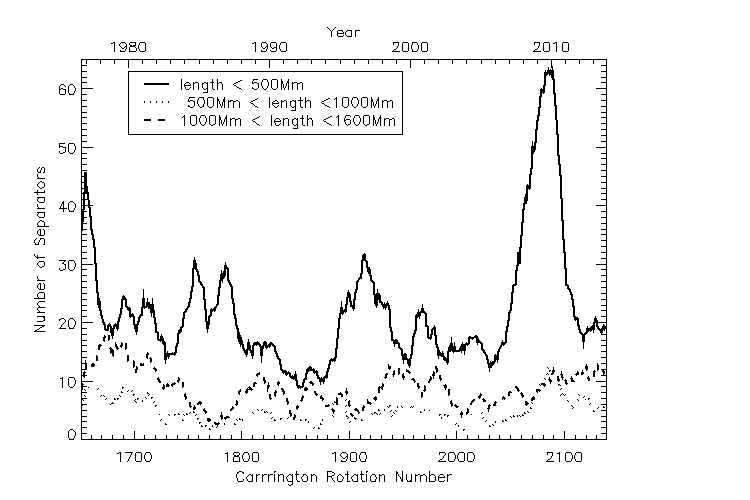}
\caption{The 12 Carrington rotation running mean time variation of the number of separators of lengths shorter than 500Mm (solid line),  between 500Mm and 1000Mm (dotted line) and between 100Mm and 1600Mm, which is the maximum length found in our model (dashed line).}
\label{sep_length}
\end{figure}
Reconnection can occur along the length of a separator (e.g., \citeauthor{galsgaard00b} \citeyear{galsgaard00b}, \citeauthor{parnell10a} \citeyear{parnell10a}) so all the separators could be sites associated with significant amounts of reconnection. Figure~\ref{sep_length} shows the variation in the length of separators over the solar cycle. Separators shorter than 500 Mm are the most common type with more of these separators found during solar minimum than solar maximum. There is no clear trend in the number of separators between 500 Mm and 1000 Mm (Figure~\ref{sep_length} dotted line). Many of these separators are likely to be ones that only reach to mid-heights in the corona.  The number of separators longer than 1000 Mm are likely either to connect two low lying null points, but {\bf arch up} to a high maximum height in the solar atmosphere, or to connect from a low lying null to a high null or even the neutral line at the base of the HCS. The longest separators are up to 1600 Mm long with most of these found during solar maxima, suggesting that most are null-HCS separators. However, during the recent solar minimum the number of long separators remained fairly high. This is most probably because the weak polar fields during the last minimum lead to more separatrix curtains, which form separators when they intersect the HCS curtain.

\section{Conclusions}
This paper examines the different topological properties of a PFSS model.
All the null points, separatrix surfaces, spines, separators, bald patches and associated separatrix surfaces are found in each potential coronal magnetic field extrapolated from a synoptic magnetogram. A total of 496 synoptic magnetograms are studied, one from each Carrington rotation from 17th August 1976 to 21st November 2013 (excluding four missing sets). The magnetogram data are specifically chosen because they all have the same resolution enabling a direct comparison between several solar cycles. However, they are not all taken with the same instrument. Up to CR2007, the data are from the Vacuum Telescope, NSO at Kitt Peak (which was upgraded in 1993). The data for CRs after 2007 are from SOLIS, NSO at Kitt Peak. Although the resolution of these instruments is the same their sensitivity is different, but we do not think this significantly effects our results. Specifically, the change in behaviour of the latest solar minimum in comparison to the previous two has been well documented and the subject of a number of major international conferences, so it is not a surprise we find evidence of it from our topological analysis. Furthermore, our findings suggest that the behaviour seen during the latest minimum is characteristic of a weak solar dipole field. This too has also been noted by other authors (e.g., \citeauthor{wang09} \citeyear{wang09}).  

From our analysis of the PFSS models it is possible for the main topological features found in global coronal field models to be classified. They include the well known null line at the source surface that forms the base of the heliospheric current sheet (HCS) and its associated HCS curtains which are the pairs of separatrix surfaces that extend down from the HCS null line. Another set of features are called separatrix curtains which are formed from one or more connected separatrix surfaces originating from coronal nulls and extending up to the source surface. Separatrix curtains that are closed (i.e., bounded on both sides by the HCS or forming an isolated closed loop themselves on the source surface) are associated with disconnected open field regions and are known to coincide with pseudostreamers. Open separatrix curtains, which are bounded on one or more edge by a spine, do not form additional open-field regions, but they do act as free standing walls or buttresses that could potentially be sites about which reconnection might occur. These features are a new phenomena on the source surface that have not been investigated before and so it would be interesting to see if they play a role in acceleration of the solar wind or in energetic solar events such as flares or coronal mass ejections.

The final set of global topological features are the various types of separatrix domes which are made up of one or more separatrix surface originating from coronal nulls, but never extending up to the source surface. The separatrix surfaces that form domes are fully bounded by the photosphere. Those that form separatrix caves and separatrix tunnels are partially bounded by the photosphere and partially by spines from other opposite polarity null points. The number of sections bounded by spines determines the number of entrances into the cave or tunnel system, with one opening signifying a cave and two or more a tunnel system. 

All these global topological features can be found at any time during the solar cycle, but their nature and frequency changes greatly from solar maximum to solar minimum. In particular, at solar maximum the HCS forms one or more closed curves within the source surface. These curves may be highly warped and can cross both poles and the equator. Additionally, many open and closed separatrix curtains are found during solar maximum, the latter (i.e., the pseudostreamers) giving rise to many disconnected open-field regions that often arise from small portions of active-region field in the photosphere. 

At cycle minima the topology of the magnetic field is highly dependent on the strength of the solar dipole. Strong solar dipole minima, were seen during the solar minima at the ends of cycle 21 and 22. Here the HCS forms an almost perfect equatorial band, with the HCS curtains enclosing essentially all the mixed-polarity quiet-Sun field on the photosphere typically out to $\pm 70^\circ$ latitude. Only a very small number of separatrix curtains are found and these are generally open, therefore do not form disconnected open-field regions. This means there are normally just two open-field regions during these minima originating from large polar areas of the photosphere.

The minimum at the end of cycle 23 is found to be a weak solar dipole minimum. Here the HCS wiggles across the equator fluctuating between $\pm 20^\circ$ or $\pm 30^\circ$ latitudes. The distortion of the HCS occurs due to the large number of separatrix curtains which arise because, with a weak dipolar field, low-latitude (e.g, $\pm 60^\circ$ latitude) quiet-Sun flux regions can have a global influence. Many of the separatrix curtains found during this minimum were closed and, hence, quite a number of disconnected open-field regions were found. However, unlike during solar maximum, large areas about the poles still remain open and the additional open-field regions are typically confined to low latitudes. 
 
At solar maximum many null points are found high up in the solar atmosphere. These nulls can have separatrix domes that enclose large amounts of flux, but above them the field is open. Thus, these nulls could be potential sites of CMEs resulting from ``breakout'' (\citeauthor{antiochos_breakout}, \citeyear{antiochos_breakout}). However, we note that many more complex separatrix domes, separatrix caves and separatrix tunnels exist. These consist of several coronal nulls connected by separators and are also important locations for reconnection. Therefore, they too are potential sites of CMEs or flares. 

Long separators, which are more numerous at maximum than at minimum, also occur. Such long separators typically link low-altitude null points with the neutral line at the base of the HCS, or connect low-altitude nulls to high-altitude nulls. These are also likely reconnection sites where magnetic energy can be released high in the corona and magnetic flux is transferred between closed and open-field regions. Further work will be undertaken to determine the nature of the reconnection and the resulting reconnected field structures in each of these cases and compared with observed flare and CME structures.

Finally we remark that, even with a simple potential field extrapolated from some of the lowest resolution synoptic magnetograms available, the magnetic field topology is highly complex both at small scales (particularly at solar minimum) and at large scales (particularly at solar maximum). However, it is important to establish how this complexity increases with, for instance, the use of high resolution SOLIS data or HMI synoptic magnetograms. So another element of our future work will be to examine the effects of changing resolution on the global topology.

\begin{acknowledgements} SJP acknowledges the financial support of the Isle of Man Government. ERP is grateful to the Leverhulme Trust for his emeritus fellowship. The research leading to these results has received funding from the European Commission's Seventh Framework Programme (FP7/2007-2013) under the grant agreement SWIFF (project no.263340, www.swiff.eu).
\end{acknowledgements}




\end{document}